\documentclass[11pt]{article}

\usepackage{amssymb}
\usepackage{theorem}
\usepackage{graphicx,epstopdf}
\usepackage{psfrag}
\usepackage{amsmath}
\usepackage{latexsym}
\usepackage{multirow}
\usepackage{subfigure}
\usepackage{lscape} 

\usepackage{filecontents} 

\usepackage[]{natbib}
\usepackage{multirow}
\usepackage{verbatim}
\usepackage{setspace}
\usepackage{subfigure}
\usepackage{float}
\restylefloat{table}

\newcommand{\boldtheta}{\mbox{\boldmath $\theta$}}

\newcommand{\boldeta}{\mbox{\boldmath $\eta$}}
\newcommand{\boldbeta}{\mbox{\boldmath $\beta$}}

\newcommand{\boldmu}{\mbox{\boldmath $\mu$}}

\newcommand{\boldomega}{\mbox{\boldmath $\omega$}}

\newcommand{\bolddelta}{\mbox{\boldmath $\delta$}}

\newcommand{\boldeps}{\mbox{\boldmath $\epsilon$}}

\newcommand{\sigeps}{\mbox{{$\sigma^2_{\epsilon}$}}}

\newcommand{\sigbeta}{\mbox{{$\sigma^2_{\beta}$}}}

\newcommand{\sigeta}{\mbox{{$\sigma^2_{\eta}$}}}

\newcommand{\boldX}{{\bf X}}

\newcommand{\boldx}{{\bf x}}

\newcommand{\boldy}{{\bf y}}

\newcommand{\boldI}{{\bf I}}

\newcommand{\boldone}{{\bf 1}}

\newcommand{\boldzero}{{\bf 0}}

\newcommand{\bc}{\begin{center}}
\newcommand{\ec}{\end{center}}
\newcommand{\bi}{\begin{itemize}}
\newcommand{\ei}{\end{itemize}}
\newcommand{\be}{\begin{enumerate}}
\newcommand{\ee}{\end{enumerate}}
\newcommand{\bd}{\begin{description}}
\newcommand{\ed}{\end{description}}
\newcommand{\bs}{\begin{single}}
\newcommand{\es}{\end{single}}

\usepackage{authblk}

\title{\Large \bf Bayesian Gaussian models for interpolating large-dimensional data at misaligned areal units}
\author[1,2]{K. Shuvo Bakar\thanks{Corresponding author. Email: shuvo.bakar@data61.csiro.au}}
\affil[1]{Data61, CSIRO, Canberra 2601, Australia}
\affil[2]{CSR\&M, The Australian National University, Canberra 2601, Australia}

\date{September, 2017}

\begin{document}

\maketitle

\abstract
Areal level spatial data are often large, sparse and may appear with geographical shapes that are regular (e.g., grid based climate model output) or irregular (e.g., postcode). Moreover, sometimes it is important to obtain predictive inference in regular or irregular areal shapes that is misaligned with the observed spatial areal geographical boundary. For example, in a survey the respondents were asked about their postcode, however for policy making purposes, researchers are often interested to obtain information at the statistical area level 2 (SA2) -- a geographical area defined by the Australian Bureau of Statistics (ABS). This level of geography is the lowest for which the ABS outputs population projections and migration data, and is used as the basis for much of the spatial analysis undertaken in Australia. The statistical challenge is to obtain spatial prediction at the SA2s, where the SA2s may have overlapped geographical boundaries with postcodes.     

The study is motivated by a practical survey data obtained from the Australian National University (ANU) Poll. 
Here the main research question is to understand respondents' satisfaction level with the way Australia is heading. The data are observed at 1,944 postcodes among the 2,516 available postcodes across Australia, and prediction is obtained at the 2,196 SA2s. This paper develops a modelling approach to address the issue with prediction or spatial interpolation at the SA2s, where postcodes and SA2s have geographical boundaries that overlaps. 

The proposed method also explored through a grid-based simulation study, where data have been observed in a regular grid and spatial prediction has been done in a regular grid that has a misaligned geographical boundary with the first regular grid-set. The real-life example with ANU Poll data addresses the situation of irregular geographical boundaries that are misaligned, i.e., model fitted with postcode data and hence obtained prediction at the SA2.
A comparison study is also performed to validate the proposed method. 

In this paper, a Gaussian model is constructed under Bayesian hierarchy. The novelty lies in the development of the basis function that can address spatial sparsity and localised spatial structure. It can also address the large-dimensional spatial data modelling problem by constructing knot based reduced-dimensional basis functions.

\noindent {\bf Keywords:} Bayesian spatial models, basis functions, misaligned areal units.

\section{INTRODUCTION}

Data collected at a defined geographic boundary is termed as the areal data in geo-statistics~\citep{banerjee2014hierarchical}. For example, infant mortality data measured at a postcode or temperature data measured in a particular grid generated from a climate model (e.g., general circulation models). It is often important to understand the spatial prediction of the variable of interest in areal positions that is geographically different than the observed areal geography. For example, data have been measured at postcodes, however spatial prediction is required at the statistical area level 2 (SA2), which is a geographically defined area in Australia that has an overlapped geographical boundary with postcode. 
The existing methods for modelling areal data with misalignment problems are mainly focused on modelling spatial data from different areal units or combining data from point-referenced and areal spatial units~\citep{bakar2015spatiodynamic,bakaretal2016,gelfand2001change}, which is also often known as the change-of-support problem~\citep{cressieandwikle11}. This way of viewing the spatial misalignment relates to analyse the model with different misaligned units instead of providing spatial prediction at the misaligned areal unit. 
Current paper aims to analyse sparse areal data and hence to predict or interpolate the variable of interest in misaligned areal units. 
Hence, this paper views the problem in a completely different way where the main intention is spatial prediction. 

The Bayesian conditional autoregressive (CAR) models~\citep{banerjee2014hierarchical} are popular to analyse areal data, which incorporates the areal spatial correlation through spatial process. 
For example, risk of having a particular disease has been often spatially correlated and inclusion of a spatially correlated process in the model provides benefits in addition to the covariates or predictor variables that may sometimes explain the spatial variations~\citep{lee2014bayesian}. This is also true for climatic data measured at the grid level, where it exhibits a strong spatial correlation~\citep{crimp2015bayesian}.
In both situations, the variable of interest is measured at the areal level (i.e., for disease risk it may be a postcode and for climatic variable -- a defined regular grid) and the Bayesian CAR models, a particular version of the Gaussian Markov random field (GMRF) can be used for estimating model parameters and also predicting at the areal spatial scale~\citep{cressieandwikle11}. However, it is often challenging to obtain spatial predictive inference at the misaligned or geographically overlapped areal units based on the existing CAR modelling strategies. For example, in this situation, the spatial process used for analysing the model is different than the spatial process used for obtaining prediction. This is due to the different layouts of the spatial geographical boundaries used in model analysis and in prediction, and therefore initiates to develop a novel modelling method to address the issue.  

The rest of the paper is organised as follows: Section~\ref{model} provides the development of the modelling technique. A simulation study is presented in Section~\ref{sim}, to test the applicability and comparative performance of the proposed technique. Section~\ref{application} describes an application of the method to the Australian National University (ANU) Poll survey data. Finally, Section~\ref{con} provides conclusions.

\section{METHODS}\label{model}

This paper develops a method based on spatial basis functions that comprises properties of both the GMRF and the Gaussian spatial random field. In particular, the Moran’s I basis~\citep{hughes2013dimension} and a Gaussian spatial basis functions (e.g., bi-square, radial) are used. The Moran’s I basis function incorporates so many properties that links with the GMRF, where the areal unit (e.g., grid, pixel, postcode, state) accounts for spatial dependence through their neighbours. For example, the conditional probability of a signal in an areal unit of a Markov random field depends on the signals in the neighbourhood through some weight neighbourhood matrix. The neighbourhood matrix usually contains a binary structure depending on the boundaries shared by the areal units, i.e., if two or more areal units share a common boundary then they are treated as one otherwise zero~\citep{lee2014bayesian}. By contrast, the spatial Gaussian basis can be constructed by calculating the distances between two spatial points, which is often more appropriate for modelling point-referenced data~\citep{bakar2017spreview}.  
In this paper, the Moran's I and bi-square basis functions are used to reflect a hybrid spatial basis, which incorporates both GRMF and Gaussian random field to connect geographical boundaries for obtaining predictions at misaligned areal units.

Figure 1 provides a conceptual flow-diagram of the modelling process implemented in this paper, where two areal layers: layer-1 $(6\times 6)$ and layer-2 $(4\times 4)$ are defined. Suppose, layer-1 data is used to fit the model with covariate information and then predict in layer-2 through the latent process (or layer) defined in the modelling hierarchy. Note that layer-1 data may be sparse or may follow a localised clustering structure, which  can be addressed by the proposed modelling approach. 

\begin{figure}[htb]
\begin{center}
\includegraphics[width=11cm, height=6cm]{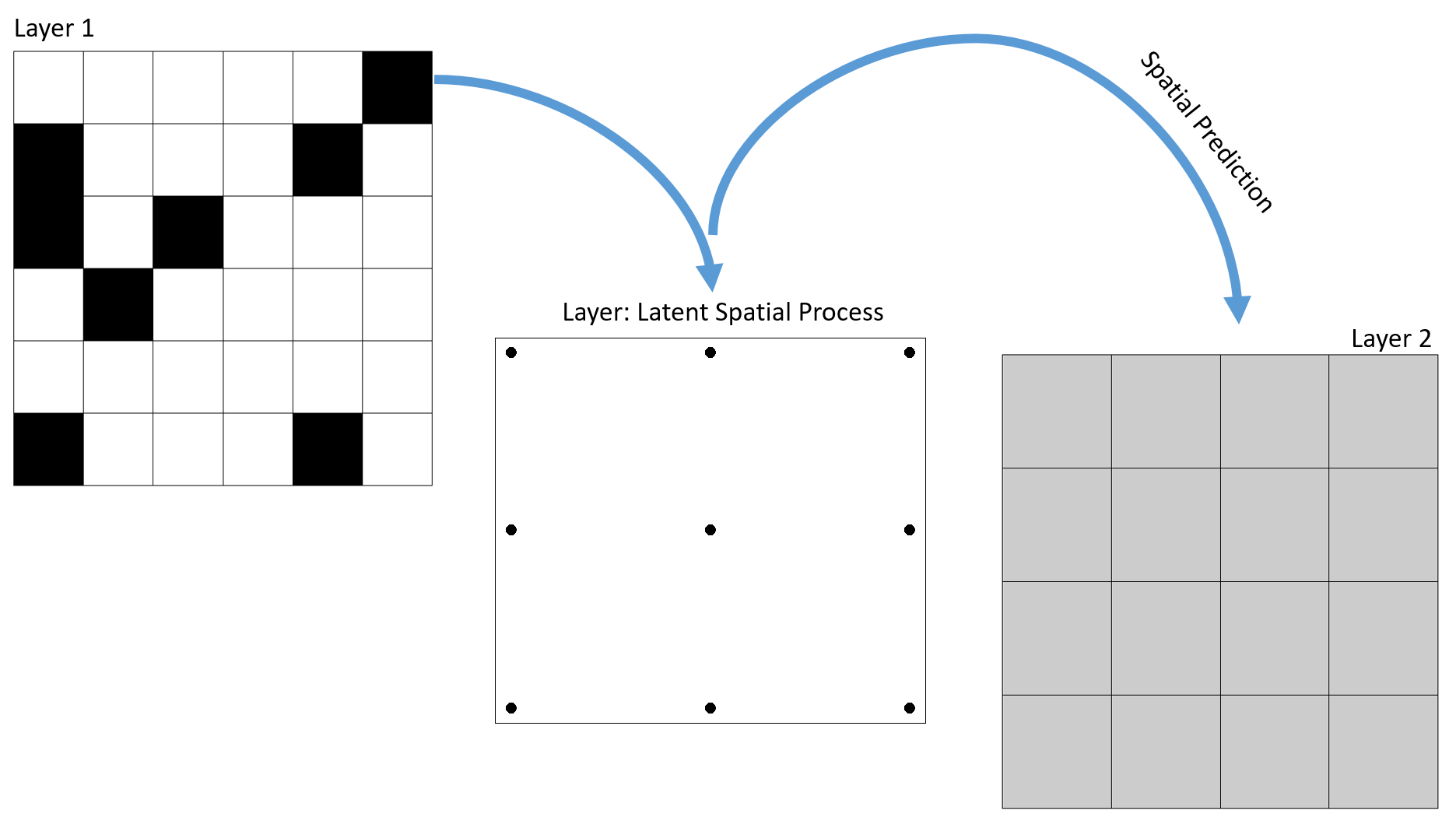}
\caption{Conceptual flow-diagram of the model using grid-based geographical layers. The layer 1 ($6\times 6$ grid) and 2 ($4\times 4$ grid) geographical boundaries overlap, and the latent spatial process (e.g., a $2\times 2$ grid) is used to link between the two layers.}
\end{center}
\end{figure}

\subsection{MODEL DEVELOPMENT}\label{dev}

Let us define $y(B_i)$ be the continuous response from a survey at areal unit $B_i$, $i=1,\ldots,n$. The aim is to obtain spatial prediction of the response $y(B^{*}_k)$ at areal unit $B^{*}_k$, $k=1,\ldots,n^{*}$, where $B_i$ and $B_k$ are spatially misaligned. Note that as discussed earlier, unlike the existing literature~\citep{banerjee2014hierarchical,cressieandwikle11}, where the areal misalignment problem is addressed at the modelling stage, this paper initiates to address it at the spatial prediction stage. The spatial model can be written as:
\begin{eqnarray}
\label{eq:1} \boldy = \boldmu+\boldeps = \boldX\boldbeta+\boldomega+\boldeps, \quad \boldeps \sim N(\boldzero,\sigma^2 \boldI)
\end{eqnarray}
where, $\boldmu=(\mu(B_i),\ldots,\mu(B_n))'$ is the mean and $\sigma^2$ is the variance of the white-noise $\boldeps=(\epsilon(B_1),\ldots,\epsilon(B_n))'$. The term $\boldX$ is an $n\times p$ design matrix, $\boldbeta$ is a $p\times 1$ parameter vector, and $\boldomega=(\omega (B_1),\ldots,\omega (B_n))'$ is the spatially correlated areal process. The CAR model is a popular choice for the spatial areal process $\boldomega\sim N(\boldzero,\tau^2 S^{-1})$, which is a particular form of the GMRF~\citep{banerjee2014hierarchical}. Here, $S$ is the precision matrix, contains a non-negative symmetric $n\times n$ neighbourhood matrix $W$ that controls the spatial autocorrelation structure of the spatial process. The $W$ matrix includes zero and one binary structure based on geographical contiguity. The CAR model is commonly specified as a set of $n$ univariate full conditional distributions $\pi(\omega_i|\boldomega_{-i})$, $i=1,\ldots,n$ where $\boldomega_{-i}=(\omega(B_1),\ldots,\omega(B_{i-1}),\omega(B_{i+1}),\ldots,\omega(B_n))'$.  
It is important to note that a discretely indexed spatial approach (e.g., CAR model) for the spatial process $\boldomega$ may exhibit smoothing variations due to the selection of the neighbourhood weight matrix $W$. Here a geographical region with small geographical boundary size is treated as the same as an area with larger geographical boundary. A localised adaptive approach of choosing the $W$ matrix can aid this problem~\citep{lee2014bayesian}, however predicting the localised spatial effect at misaligned areal units (e.g., $B^{*}_k$, $k=1,\ldots,n^{*}$) with overlapping geographic boundaries is often challenging. 

The use of the Moran's I basis function is beneficial in this context to represent the discretely indexed spatial random field, which is aimed to solve the collinearity between the fixed effect estimates and the CAR process~\citep{hughes2013dimension,reich2006effects}. This way of defining the correlated process initiates sampling from a multivariate full conditional distribution of $\boldomega$. In the situation of a large number of areal units, the problem arises on inverting the large $n\times n$ spatial covariance matrix with $\mathcal{O}(n^3)$ computational complexity. Importantly, the Moran's I approach can aid the large spatial data problem through dimension reduction~\citep{bradley2016bayesian}, where a smaller number of knot points $r(<n)$ is considered in the discrete spatial random field to represent the reduced-dimension. Similarly, a reduced dimensional method (e.g., fixed rank Kriging) can be applied in the continuous Gaussian spatial field using spatial basis functions such as wavelet, radial, bi-square~\citep{cressieandjohannesson08,nychka2015multiresolution}. Here, the knot points are defined in the continuous Gaussian spatial field. 

This paper defines the spatially correlated process $\boldomega$ as a function of the Moran's I and bi-square basis functions. Hence, the spatial process $\boldomega$ is written as:
\begin{eqnarray}
\label{eq:2} \boldomega = \Lambda\bolddelta + \boldeta, \quad \boldeta\sim N(\boldzero,\sigeta\boldI),
\end{eqnarray}
where, $\Lambda=M\times R$ is a basis function that includes an $n\times n$ matrix of eigenvectors $M$ of the Moran's I operator matrix $M(W)$, and the lower-dimensional $n\times r$ bi-square basis function $R$. 
Following~\cite{bradley2016bayesian}, we write the Moran's I operator matrix as:
\begin{eqnarray}
\label{eq:3} M(W) = (\boldI-\boldX(\boldX'\boldX)^{-1}\boldX')W(\boldI-\boldX(\boldX'\boldX)^{-1}\boldX')
\end{eqnarray}
where, $\boldI$ is an $n\times n$ identity matrix, and $W$ is the $n\times n$ nearest neighbour weight matrix contains zeros and ones. The term $\boldeta\sim N(\boldzero,\sigeta\boldI)$ is an i.i.d. error process that captures the remaining random component.~\citet{hughes2013dimension} reduced the Moran's I basis function into a lower dimension for achieving the spatial dimension reduction. However, following~\citet{reich2006effects} this paper includes the full dimension (i.e., $n\times n$) of the spectral decomposition of $M(W)$ and proposes the reduction of spatial dimension using the bi-square basis functions~\citep{cressieandjohannesson08}. Hence, $r(<n)$ knot points are defined to design the $n\times r$ bi-square basis matrix $R$.
The basis $R(B_{ij})$, $i=1,\ldots,n$, $j=1,\ldots,r$ is constructed using the distances between the centroid of the areas and the knot points. Thus the $n\times r$ spatial basis $\Lambda$ includes both discrete neighbourhood phenomenon of the geographical areas (i.e., $M$) and their distances through bi-square basis (i.e., $R$). 
It is always challenging to select the number and position of the knots for defining $R$~\citep{sahuandbakar2012b}. This paper considers a regular grid for selecting the knot positions and the number of knot points $r$ is selected based on a compromising choice between the improved accuracy and the computational efficiency of the model~\citep{sahu2015bayesian}. 

The lower dimensional vector process $\bolddelta$ defined in eq.~(\ref{eq:2}) is assumed to have a zero mean and lower dimensional $r\times r$ covariance matrix $\Sigma$. A smoothing parameter $\phi$ is used to define $\Sigma$ and is written as: $\Sigma=\phi\times \Lambda' Q \Lambda$. The term $Q$ is defined as $Q=\text{diag}(W\boldone) - W$, where $\boldone$ is a vector of ones. 
With a spectral decomposition of $\Lambda' Q \Lambda$, we write $\Sigma=\Psi_Q(\phi\times \Phi_Q)\Psi_Q'$, where $\Psi_Q$ is an orthogonal matrix of order $r\times r$ and $\Phi_Q$ is a diagonal matrix.~\citet{bradley2016bayesian} used a prior distribution on $\Psi_Q$; however, this paper assumes that the eigenvectors of $\Sigma$ and its eigenvalues are known up to a multiplicative constant~\citep{hughes2013dimension}. 

\subsection{Posterior}
The joint posterior distribution of the data, processes and parameters of the proposed model is represented as a product of the following conditional distributions:
\begin{align}
\nonumber &\text{Data Model: } \boldy\left|\boldmu,\sigma^2 \right.\sim \text{N}\left(\boldmu,\sigma^2\boldI_{n\times n}\right), \\
\nonumber &\text{Process Models: } \boldmu\left|\phi \right.\sim \text{Gaussian}\left(\boldx\boldbeta+\Lambda\bolddelta,\sigeta\boldI_{n\times n}\right), \bolddelta\left|\phi \right.\sim \text{Gaussian}(\boldzero,\Sigma),\\
\nonumber &\text{Parameter Models: } \boldbeta \sim \text{Gaussian}\left(\boldmu_\beta,\sigbeta\boldI_{p\times p}\right), \sigma^2 \sim \text{Inverse-Gamma}(a_{\epsilon},b_{\epsilon}), \\ 
\nonumber &\quad\quad\quad\quad\quad\quad\quad\quad \sigeta \sim \text{Inverse-Gamma}(a_{\eta},b_{\eta}),\phi \sim \text{Inverse-Gamma}(a_{\phi},b_{\phi}),
\end{align}
where, $\sigma^2>0$, $\sigeps>0$ and $\phi>0$. We use proper and non-informative prior distributions for $\beta$ by setting $\mu_\beta=0$ and $\sigbeta=10^{10}$. For the hyper-parameters of the inverse-gamma prior distributions, we set $a_{\epsilon}=a_{\eta}=a_{\phi}=2$ and $b_{\epsilon}=b_{\eta}=b_{\phi}=1$, as the $\text{Inverse-Gamma}(2, 1)$. 
The Markov chain Monte Carlo (MCMC) sampling algorithm is used with the {\tt R} programming language and the just another Gibbs sampler (JAGS)~\citep{plummer2003} computational platform to obtain estimates of the model parameters and hence spatial predictions. Detail code are provided as a supplementary material.  

\subsection{Prediction}

Let us define a geographical area $B_k^{*}$, $k=1,\ldots,n^{*}$ where we want to obtain prediction that has overlapped geographical boundary with the area where we observe data $B_i$, $i=1,\ldots,n$. The predictive posterior distribution for $y(B_k^{*})$ at $B_k^{*}$ can be obtained by solving the following integral.
\begin{eqnarray}\label{eq:7}
\nonumber \pi\left(y\left(B_k^{*}\right)\left|\boldX^{*},\Lambda^{*}\right.\right) &=& \int \pi\left(y\left(B_k^{*}\right)\left|\boldmu^{*},\boldtheta\right.\right) \times \pi\left(\boldmu^{*}\left|\boldX^{*},\Lambda^{*},\boldtheta\right.\right) \times \pi\left(\boldtheta\left|\boldx^{*}\right.\right) \text{d}\boldmu^{*}\text{d}\boldtheta; 
\end{eqnarray}
where, $\boldX^{*}$ is the design matrix at $B_k^{*}$. We define the term $\boldtheta=(\boldbeta,\bolddelta,\phi,\sigma^2,\sigeps)'$ and the term $\Lambda^{*}=M^{*}\times R^{*}$. Note that the neighbourhood weight matrix $W^{*}$ is calculated from $B_k^{*}$, $k=1,\ldots,n^{*}$. 

\section{SIMULATION: A COMPARATIVE STUDY}\label{sim}

A simulation study is used to demonstrate the sensitivity and predictive performance of the proposed model. The model is compared with the Moran's I basis function model~\citep{hughes2013dimension}. Regular and overlapping grids are used in this context. The knot sizes are also assumed to be same for both models. For example, it is often ideal to use 10\%~\citep{hughes2013dimension,bradley2016bayesian} of the total number of spatial units $(n)$ as the number of knots $(r)$ for reducing the dimension for Moran's I method. Hence, 10\% of the $n$ is considered as $r$ in this paper for defining the knot size for the bi-square basis function. A sensitivity analysis on different knot sizes is also performed in this context.

Two sets of $30\times 30$ and $20\times 20$ overlapped grids are considered, where $30\times 30$ is used to fit the model and then predicted in $20\times 20$. A total $r=85$ knot points (about 10 percent of the 900 grids) are used for constructing the spatial basis function. 
Let us denote the $30\times 30$ as lattice--1 and $20\times 20$ as lattice--2. Two predictor variables are simulated from a standard normal distribution, and the response variable is constructed correspondingly. Note that lattice--1 is used for model fitting and model based spatial predictions are obtained in lattice--2, that has an overlapped boundary with lattice--1. The true simulated population measurements are stored to compare the model output in lattice--2. 

Comparison of the proposed and the Moran's I model are then obtained based on their predictive performance. The root mean squared error (RMSE) and deviance information criterion (DIC)~\citep{spiegelhalter2002bayesian} are used as the model comparison criteria. The proposed model shows a lower DIC value (1780.81) and RMSE estimate (0.27) compared to the Moran's I model (DIC=1957.01, RMSE=0.35). Based on the summary statistics it can be concluded that the proposed model performs better than the Moran's I model for predicting at the misaligned geographical boundaries. 
\begin{table}[htp]\caption{\label{tab:val} The validation criteria: root mean squared error (RMSE) with 95\% Bayesian crediable interval, and deviance information criterion (DIC) with measure of fit and corresponding penalty.}
\centering
\begin{tabular}{lrr} \hline
  & RMSE (95\% Bayesian--CI) &  DIC (measure of fit + penalty) \\ \hline
Moran's I model & 0.3501 (0.2852,0.4397) & 1957.01 (874.65+1082.36)\\
Proposed model & 0.2712 (0.2150,0.3652) & 1780.81 (593.24+1187.57)\\ \hline
\end{tabular}
\end{table}

\section{APPLICATION: ANU POLL DATA ANALYSIS}\label{application}

\begin{figure}[htb]
\begin{center}
               \makebox{\subfigure[POA]{\includegraphics[width=0.23\textwidth]{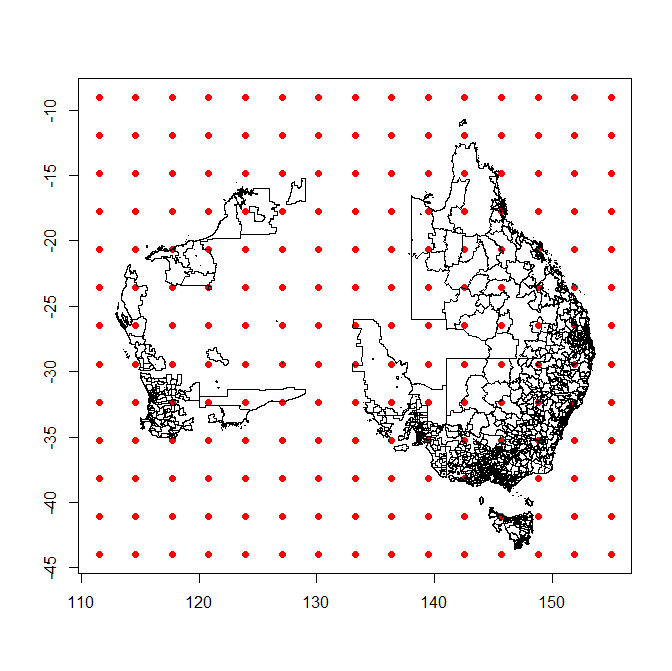}}}
               \makebox{\subfigure[SA2]{\includegraphics[width=0.23\textwidth]{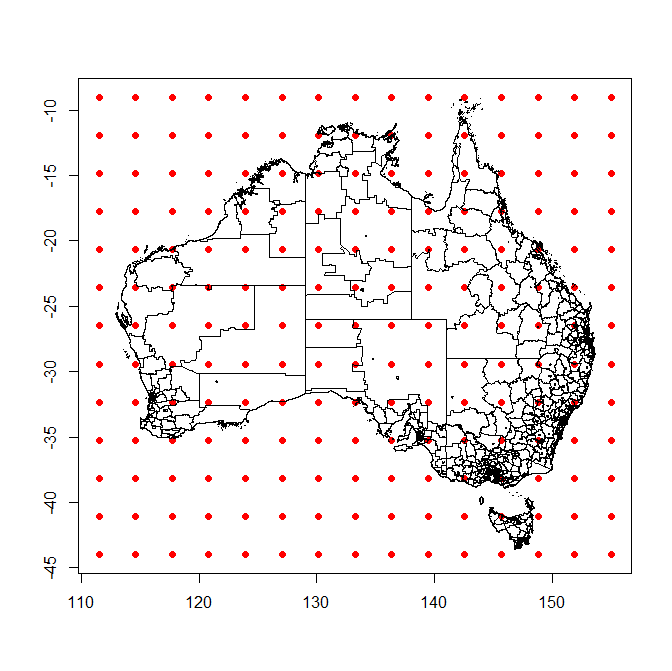}}}
               \makebox{\subfigure[SA2]{\includegraphics[width=0.23\textwidth]{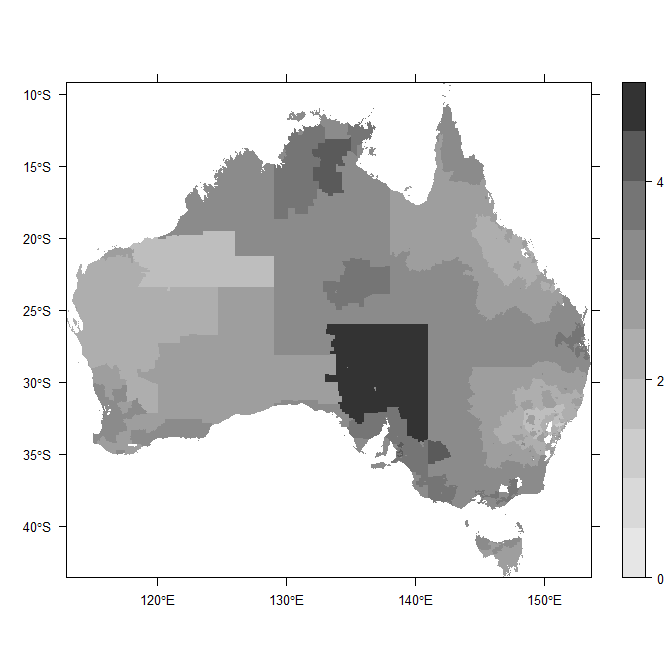}}}
               \makebox{\subfigure[SA2]{\includegraphics[width=0.23\textwidth]{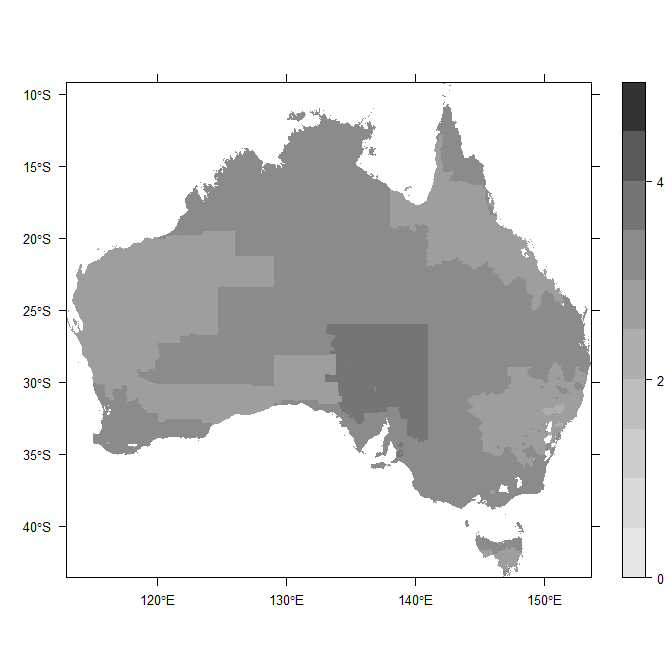}}}
               \makebox{\subfigure[Brisbane]{\includegraphics[width=0.23\textwidth]{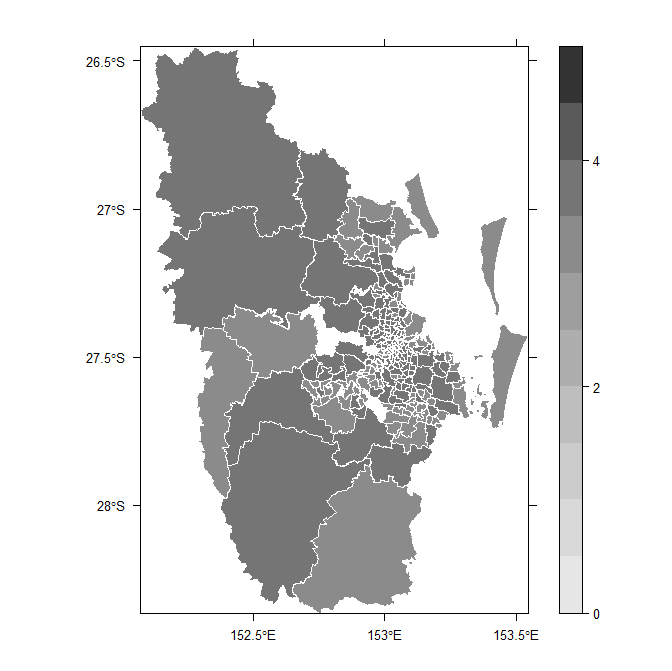}}}
               \makebox{\subfigure[Sydney]{\includegraphics[width=0.23\textwidth]{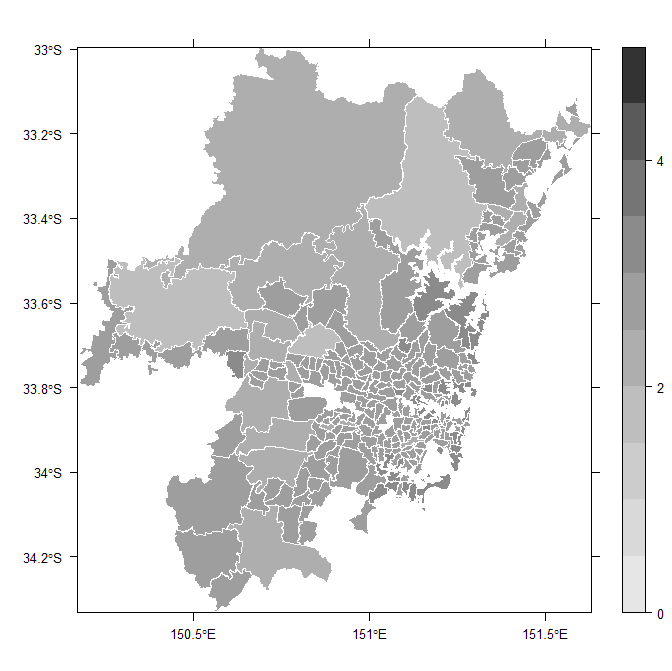}}}
               \makebox{\subfigure[Melbourne]{\includegraphics[width=0.23\textwidth]{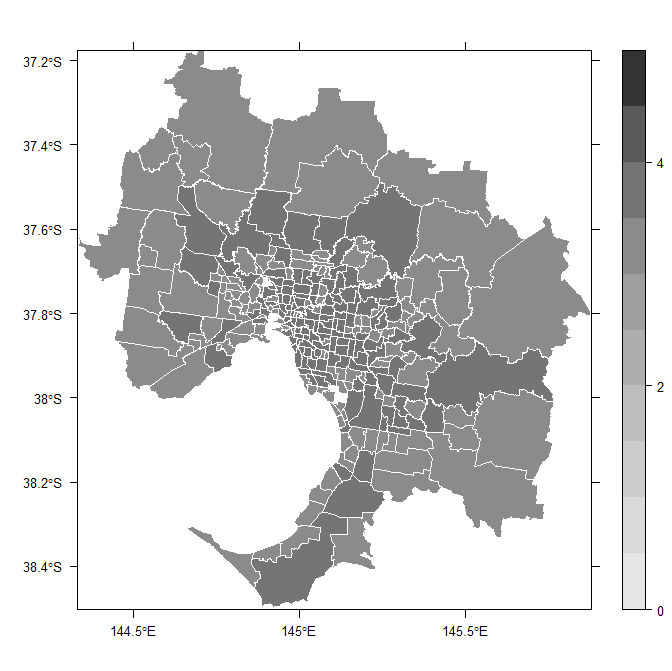}}}
               \makebox{\subfigure[Adelaide]{\includegraphics[width=0.23\textwidth]{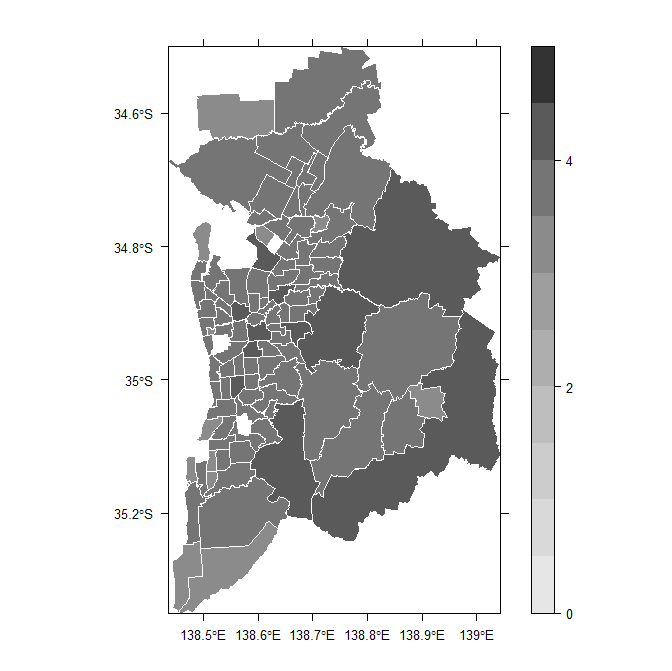}}}
               \caption{\label{fig:map} (a) Total 1,944 Postcodes in the ANU Poll data and (b) 2,196 postcodes across Australia, with 195 points used for defining the knots. (c) Model based predictive map of Australia at the SA2 level, (d) prediction without adjusting spatial effects in the model. Model based predictive map at the SA2 level for Greater (a) Brisbane (b) Sydney (c) Melbourne and (d) Adelaide.}
\end{center}
\end{figure}

The ANU Poll has been a nation-wide survey conducted three times each year. A pooled ANU Poll dataset from June 2010 to October 2015 is used in this application. A Total of 13 surveys was conducted in this period. One of the common questions in the surveys was on respondents' satisfaction level with the way Australia is heading. A total of 18,978 respondents was recorded at 1,944 postcodes in the pooled data among the 2,516 available postcodes across Australia, see Figure~\ref{fig:map}(a). The responses are then aggregated at the postcode level. Although the individual survey data are in the Likert scale~\citep{likert1932technique}, i.e., 1 as very less satisfied and 5 as the most satisfied; the aggregated average response at the postcodes refers to a bell--shape normal distribution. 
The aggregated data are used for model fitting and then the spatial prediction of the average satisfaction level is obtained at the SA2s -- a spatially overlapped and misaligned geographical area with postcode. Figure~\ref{fig:map}(b) represents 2,196 SA2s across Australia. 

The socio-economic status of an area sometimes plays a key role to motivate the respondents' satisfaction. Hence, two SEIFA (Socio-Economic Indices for Areas) indices are used as predictors in the model: (1) IRSD (index of relative socio-economic disadvantage) and (2) IEO (index of education and occupation)\footnote{http://www.abs.gov.au/ausstats/abs@.nsf/mf/2033.0.55.001}. The population size (in logarithmic scale) at the areal unit is also considered as a predictor variable in the model. 

Ten percent (i.e., $r=195$) of the total postcodes (i.e., 1,944) is considered as the number of knot points (see Figure~\ref{fig:map}). The knot positions are considered as grids. 
The MCMC summary statistics show that the predictor variable IRSD does not have a significant statistical association with respondents' satisfaction with the way the country is heading. However, IEO and population size in each geographical area exhibit a positive and statistically significant effect\footnote{Bayesian--CI: at 95\% Bayesian crediable interval}. The variance parameters are estimated at 0.18 and 0.27 for the error process $\boldeps$ and $\boldeta$ respectively. The spatial smoothing parameter $\phi$ is estimated 1.02. 

Figures~\ref{fig:map}(c) and (d) show the model based predictions at the SA2s with and without spatial effects respectively. Here, the spatial modelling contributes strongly to identify geographical variations over Australia. For example, the Outback (SA2:406021141) in South Australia yields a higher average satisfaction level compared to the prediction without adjusting the spatial effects in the model. Here, Outback -- a large SA2 overlaps with approximately three large rural postcodes 5440, 5710 and 5731 in South Australia. Hence, prediction at Outback using the proposed modelling technique captures the spatial variability much higher, which was not addressed only by the predictor variables used in the model. 
Furthermore, the predictive maps are elaborated into small scale for four major capital cities in Australia in Figures~\ref{fig:map}(e)--(h). The predicted average response of Greater Adelaide shows a higher tendency of satisfaction, however Greater Sydney shows an overall lower level of satisfaction compared to other capital cities in Australia. 


\section{CONCLUSIONS}\label{con}

This paper presents a Bayesian spatial Gaussian model to enable the prediction of areal data at overlapped and misaligned geographical boundaries through a non-stationary hybrid spatial basis function. The hybrid spatial basis combines the Moran's I and bi-square basis functions. The approach addresses the big-$n$ spatial problem by considering a set of lower dimensional knots. The proposed method however, does not aim to address the problem of confounding for spatial random processes, which is left as an extension.

The method is validated using a simulation study, where the applicability is tested using regular grid datasets. The proposed model outperforms the Moran's I basis function model. 
The model is then applied to the ANU Poll survey data, where postcode information is used to obtain spatial predictive inference at the SA2 to understand the distribution of respondents' views towards the way Australia is heading. 

The proposed model can be extended to accommodate non-Gaussian models and for addressing spatio-temporal data. It is often important in many research situations to understand the effects of some key variables among a vast list of predictors~\citep{george1993variable,tibshirani1996regression}. Hence, the proposed model can be extended to address this problem.

\section*{Acknowledgement}
The author would like to thank Professor Sujit Sahu, University of Southampton for his comments on a draft of this paper. Author also wish to acknowledge the support of the CSR\&M, ANU.


\bibliographystyle{chicago} 
\bibliography{\jobname}

\end{document}